\documentclass[prl,secnumarabic,amssymb, nobibnotes, aps, prd,aps,twocolumn]{revtex4-1}
\usepackage{graphicx , epsfig,color}
\usepackage{amsmath}
\usepackage{latexsym}
\usepackage{amstext}
\usepackage{array}
\usepackage{multirow}
\usepackage{changebar}
\usepackage{subfigure}
\usepackage{stackrel}
\usepackage{ulem}

\newcommand{\mb}[1]{ \mbox{\boldmath$#1$} }
\newcommand{\ds}{\displaystyle}
\newcommand{\beq}{\begin{eqnarray}}
\newcommand{\eeq}{\end{eqnarray}}
\newcommand{\beqq}{\begin{eqnarray*}}
\newcommand{\eeqq}{\end{eqnarray*}}
\newcommand{\p}{\partial}

\newcommand{\eps}{\varepsilon}
\newcommand{\w}{\mbox{\boldmath$w$}}
\newcommand{\Tt}{\mbox{\boldmath$t$}}
\newcommand{\x}{\mbox{\boldmath$x$}}
\newcommand{\n}{\mbox{\boldmath$n$}}

\newcommand{\B}{\mbox{\boldmath$B$}}
\newcommand{\y}{\mbox{\boldmath$y$}}

\begin{document}
\title{ \textbf{Oscillatory decay of the survival probability of activated diffusion across a limit cycle }}
\author{K. Dao Duc$^{1,2}$  Z. Schuss$^3$ D. Holcman$^{1,2}$} \affiliation{$^1$Ecole Normale Sup\'erieure, $^2$Group of Applied Mathematics and Computational Biology, IBENS, 46 rue d'Ulm 75005 Paris, France\footnote{This research
is supported by an ERC-starting-Grant.} $^3$Department of Applied Mathematics,
Tel-Aviv University, Tel-Aviv, Israel.}
\date{\today}

\begin{abstract}
Activated escape of a Brownian particle from the domain of attraction of a stable
focus over a limit cycle exhibits non-Kramers behavior: it is non-Poissonian.
When the attractor is moved closer to the boundary oscillations can be discerned in the survival probability.
We show that these oscillations are due to complex-valued higher order eigenvalues of the Fokker-Planck operator, which we compute explicitly in the limit of small noise. We also show that in this limit the
period of the oscillations is the winding number of the activated stochastic process.
{These peak probability oscillations are not related to stochastic resonance and should
be detectable in planar dynamical systems with the topology described here.}
\end{abstract}
\maketitle
Thermal activation over a potential barrier consists in the escape of a noisy dynamical
system from the domain of attraction of a stable equilibrium point of the drift, as
described in Kramers' theory \cite{Kramers}. The activation process is the generic
model of many processes in physics, chemistry, tracking, and the manifestation of many
molecular and cellular processes, to mention but a few. The mean first passage time
(MFPT) of the random trajectories to the boundary of the domain is a measure of the
stochastic stability of the system under {noisy perturbations} and is determined by
the depth of the potential well. In Kramers' theory the escape process is Poissonian
for sufficiently long times, with rate that is one-half the reciprocal of the MFPT
\cite{Schuss77,Freidlin,Matkowsky,Hanggi,Gang,MaierStein1,MaierStein2,DSP}.\\
However, this is not the case for nonconservative noisy dynamics, in which the drift
is not a gradient of a potential: the escape process from the domain of attraction of a
meta-stable point is no longer Poissonian. We study here the noise-induced escape from
the domain of attraction $D$ of a stable focus of nonconservative planar dynamics
$\mb{b}(\x)$ across the boundary of $D$, which is assumed to be a repelling limit cycle. This situation arises, for example, in the damped Langevin equation in the
phase plane, in synchronization loops in communications theory, and more. The renewed
interest in this problem is due to it manifestation in models of neuronal activity \cite{Holcman2006}.
This mathematical model was considered in \cite{MaierStein1,MaierStein2} for the
steady state of a system with reinjection of escaping trajectories. It was shown that
for { noise intensity} $\eps\to0$, the arrival rate at the absorbing boundary has the asymptotic
representation $R\sim\times\eps^bG(|\log\eps|)\exp\{-\Delta \psi/\eps\}$, where $\Delta
\psi$ is the depth of the nonequilibrium potential, the parameter $b$ is
model-dependent, and the factor $G(|\log\eps|)$ is a model-dependent
periodic function of $|\log\eps|$. It is also shown that prior to absorption in the
boundary the random trajectories wind around the attractor.\\
\begin{figure}[http!]
\centering
\includegraphics[width=0.45\textwidth]{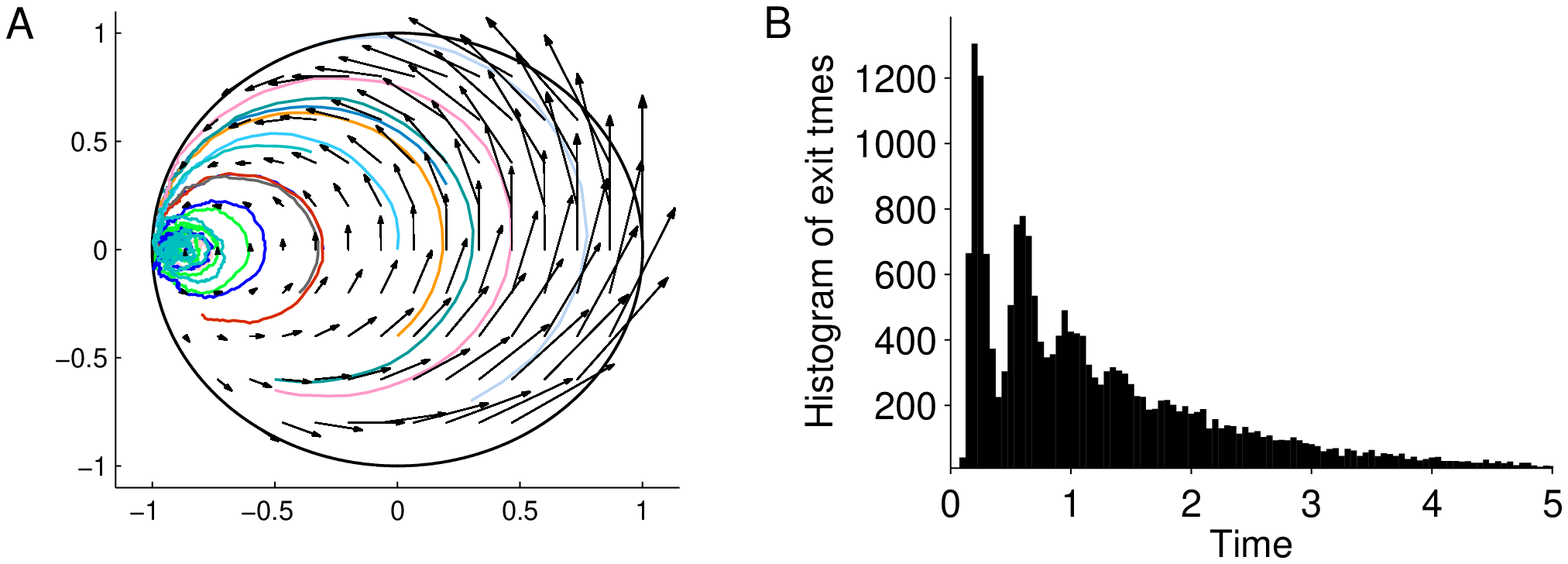}
\includegraphics[width=0.45\textwidth]{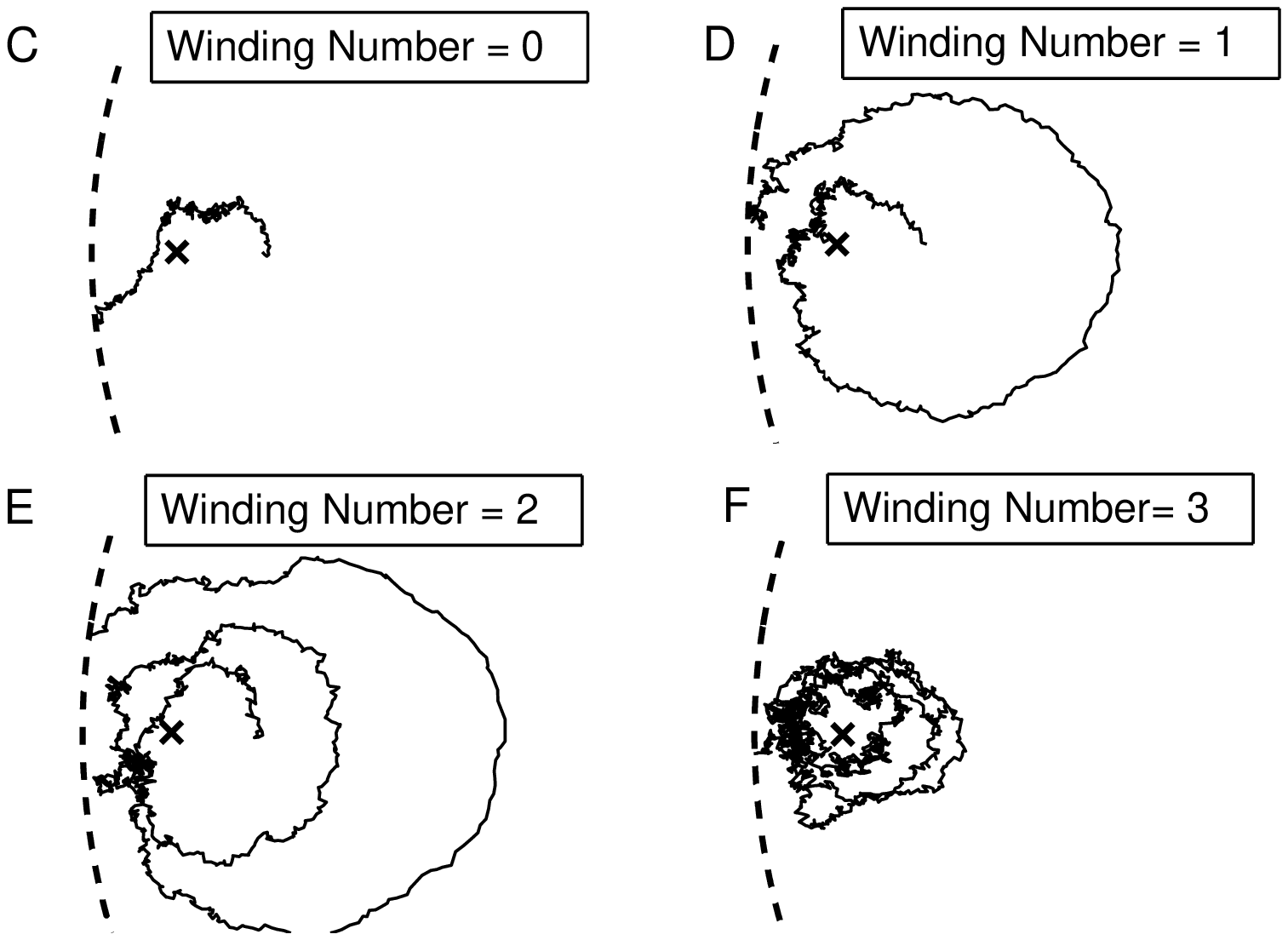}
\caption{\textbf{Trajectories and histogram of exit times. A:} Trajectories
generated by (\ref{SDEo}) with the field (\ref{field}) in the unit disk. \textbf{B:}
Histogram of exit times associated to \ref{SDEo} from the unstable limit cycle (disk). Parameters of the simulation are $\alpha=0.9$, $\varepsilon=0.001,\omega=15$,
number of simulations = 20000, initial point : (-0.5,0).}
\label{f:trajectories}
\end{figure}
We study here the time-dependent escape problem and find time-dependent oscillations of
the escape rate. When the attractor is moved toward the boundary this effect is
manifested through periodic peaks in the distribution of the exit time. These
oscillations in the survival probability, previously mistaken for a manifestation of
stochastic resonance, are shown to be the result of complex-valued higher-order eigenvalues of the
Dirichlet problem for the Fokker-Planck operator inside the limit cycle. We compute the
entire spectrum of the problem and show that the period of the oscillations is determined
by the imaginary part of the second eigenvalue. The oscillatory peaks are observable
when the real part of the second eigenvalue is comparable to the (real-valued
first eigenvalue, which is the reciprocal of the MFPT. {It is shown here that the oscillation of the exit probability
peaks is a generic phenomenon in noise-driven planar dynamical systems that have the topology of the classical Hopf system. They become discernible when the focus is moved toward the repelling boundary.} \\
{\bf Absorbing boundary and survival probability.} We consider the planar stochastic dynamics
\beq
\dot{\x}_\eps(t)=\mb{b}(\x_\eps(t))+\sqrt{2\eps}\,\mb{a}(\x_\eps(t))\dot{\w}(t),\label{SDEo}
\eeq
with a drift field $\mb{b}(\x)$ that has a stable focus $\x_0$, whose domain of
attraction $D$ is bounded by an unstable limit cycle $\p D$ (see
Fig.\ref{f:trajectories}). Here
$\dot{\w}(t)$ is $\delta$-correlated Gaussian white noise ($\w(t)$ is the Wiener
process) and $\sqrt{2\eps}\,\mb{a}(\x)$ is the diffusion matrix, scaled by a small
parameter $\eps$. The trajectories of (\ref{SDEo}) are terminated instantaneously the
moment they hit $\p D$ for the first time. It is well-known that even for
arbitrarily small $\eps>0$ the trajectories of (\ref{SDEo}) exit $D$ in finite time
with probability 1 and have finite mean, called the mean first passage time
$\bar\tau_{\eps}$ (MFPT) \cite{Schuss77,Matkowsky,Freidlin,Hanggi,DSP}.

The survival probability ${\Pr}_{\scriptsize\mbox{surv}}(t)$ of trajectories in $D$,
prior to termination at the boundary $\p D$, averaged with respect to an initial
density distribution $p_0(\x)$, can be expressed in terms of the transition
probability density function (pdf) $p_\eps(\y,t\,|\,\x)$ of the surviving trajectories
as
\begin{align}\label{surv}
{\Pr}_{\scriptsize\mbox{surv}}(t)
 =\int\limits_D\int\limits_Dp_\eps(\y,t\,|\,\x) p_0(x) \,d\y\,d\x.
 \end{align}
The pdf is the solution of the initial-boundary value problem for the Fokker-Planck
equation
\beq
\frac{\p p_\eps(\y,t\,|\,\x)}{\p t}&=&\,L_{\y} p(\y,t\,|\,\x)\hspace{0.5em}\mbox{for}\
\x,\y\in
D \nonumber\\ & & \nonumber\\
p_\eps(\y,t\,|\,\x)&=&\,0\hspace{0.5em}\mbox{for}\ \x\in\p D,\ \y\in D,\ t>0
\label{FPE}
\\ & & \nonumber\\
p_\eps(\y,0\,|\,\x)&=&\,\delta(\y-\x)\hspace{0.5em}\mbox{for}\ \x,\y\in D \nonumber
\eeq
where the Fokker-Planck operator $L_{\y}$ is given by
\begin{align*}
L_{\y}u(\y)=&\,\eps\sum_{i,j=1}^2  \frac{\p ^2\left[ \sigma ^{i,j}\left(\y\right)
u(\y) \right]}{\p y^i\p y^j}-\sum_{i=1}^2\frac {\p \left[
b^i\left(\y\right)u(\y)\right]} {\p y^i}.
\end{align*}
Here  $\mb{\sigma}(\x)=\mb{a}(\x)\mb{a}^T(\x)$. In the case at hand the operator $L_{\y}$ with the homogeneous Dirichlet
boundary conditions (\ref{FPE}) is non-self-adjoint, has complex-valued higher-order
eigenvalues $\lambda_{n,m}$, and the eigenfunctions $u_{n,m}(\y)$ of $L_{\y}$ and
$v_{n,m}(\x)$ of $L_{\x}^*$ form bi-orthonormal bases.
Only the principal eigenvalue $\lambda_0$ is positive and so are the corresponding
eigenvalues $u_0(\y)$ and $v_0(\x)$.
The general solution of Fokker-Planck initial-boundary value problem (\ref{FPE}) can be
expanded as
\beq
 p_\eps(\y,t\,|\,\x)&=& e^{- \lambda_0t}u_0(\y)v_0(\x)\label{pepsunif}\\
 &+& \sum_{n,m} e^{-\lambda_{n,m}t}u_{n,m}(\y)\bar v_{n,m}(\x).\nonumber
 \eeq
The probability density of the exit time $f_{etd}(t)$(DET) is given by
\beq
f_{etd}(t)=-\frac{d}{dt}{\Pr}_{\scriptsize\mbox{surv}}(t)
=\lambda_0e^{- \lambda_0t}+ \sum_{n,m} C_{n,m} e^{- \lambda_{n,m}t},\label{pdf}
\eeq
where $C_{m,n}$ are constants. It is the purpose of this letter to show that this density decays with large oscillations
and we present a generic case in which they are easily discernible.\\
{\bf The spectrum of the non-self-adjoint Fokker-Planck operator}\\
{\em The field $\mb{b}(\x)$.}
The local geometry of the drift field $\mb{b}(\x)$ near the focus $\x_0$ and the
unstable limit cycle $\p D$ can be described as follows. Near $\x_0$ the local behavior
of $\mb{b}(\x)$ is $\mb{b}(\x)=\B(\x-\x_0)+O(|\x-\x_0|^2)$ where the matrix $\B$ has eigenvalues in the left half of the complex plane.
The local representation of the field $\mb{b}(\x)$ in the boundary strip is given by
\beq \label{arhos}
\mb{b}_{\alpha}(\rho, s) = -\rho b^0(s)\n + B(s)\Tt
\eeq
the tangential component of the field at $\p D$ is
$B(s)=\mb{b}(0,s)\cdot\nabla s=|\mb{b}(\x(s))|>0$ and the normal derivative of the
normal component is $b^0(s)\geq0$ for all $0\leq s\leq 2\pi$. 

{\em WKB structure of the pdf for small $\eps$.}
To compute the spectrum of $L{\y}$, we use matched asymptotics \cite{Bender} to
construct a  uniform asymptotic approximation to the eigenfunctions $u_{m,n}(\y)$. We
begin with the outer expansion of and eigenfunction $u(\y)$ in the WKB form
\begin{equation}
u(\y) = K_{\ds\eps}(\y) \exp\!\left\{-\frac{\psi(\y)}\eps \right\},
\label{WKBMD1}
\end{equation}
where the eikonal function $\psi(\y)$ is a solution of eikonal equation
\begin{align}
\mb{\sigma}(\y)\nabla\psi(\y)\cdot\nabla\psi(\y)+\mb{a}(\y)\cdot\nabla\psi(\y)=0\label{eikonalBF171}
\end{align}
(see \cite[Chap.10]{DSP}) and $\psi(\y)$ is constant on the boundary.
The function $K_{\ds\eps}(\y)$ is a regular function of $\eps$ for $\y\in D$,
but has to develop a boundary layer to satisfy the homogenous Dirichlet boundary
condition $ K_{\ds\eps}(\y)=0\hspace{0.5em}\mbox{for}\ \y\in\p D.$ Therefore $K_{\ds\eps}(\y)$ is further decomposed into the product
\begin{align}
K_{\ds\eps}(\y)= \left[K_0(\y) + \eps K_1(\y) +\cdots\right]
q_{\ds\eps}(\y),\label{decomposeK}
\end{align}
where $K_0(\y),\, K_1(\y),\,\ldots$ are regular functions in $ D$ and on its
boundary and are independent of $\eps$, and $q_{\ds\eps}(\y)$ is a boundary
layer function.  The boundary layer function $q_{\ds\eps}(\y)$ satisfies the boundary
condition
$q_{\eps}(\y)=0 \mbox{ for } \y \in \p D,$ the matching condition $ \lim_{\eps\to
0}q_{\eps}(\y) =1 \mbox{ for all }  \y\in D.$

\begin{figure}[http!]
\centering
\includegraphics[width=0.2\textwidth]{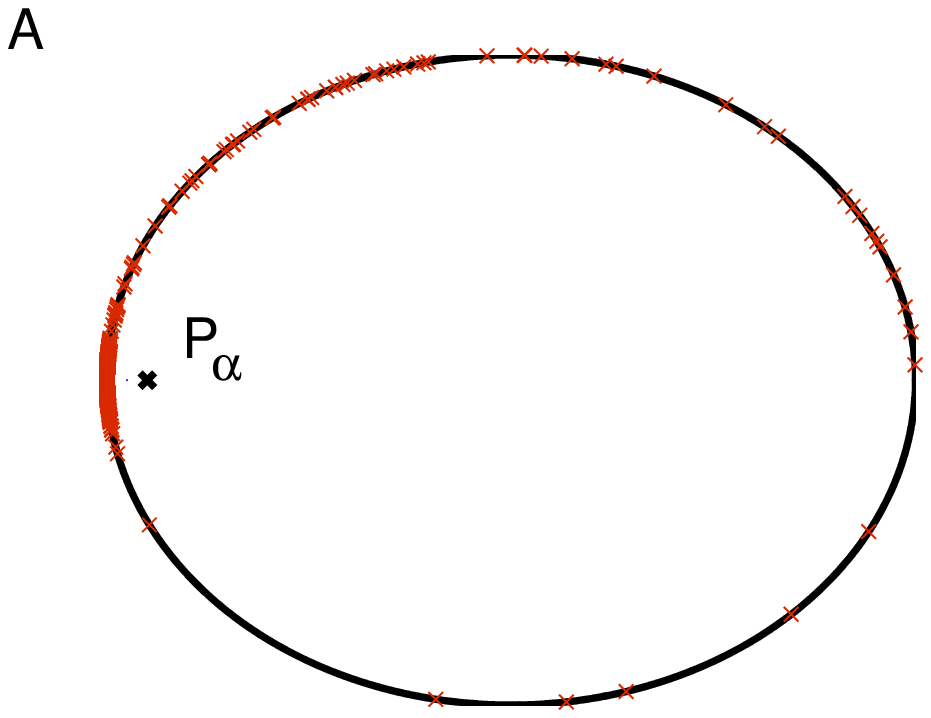}
\includegraphics[width=0.25\textwidth]{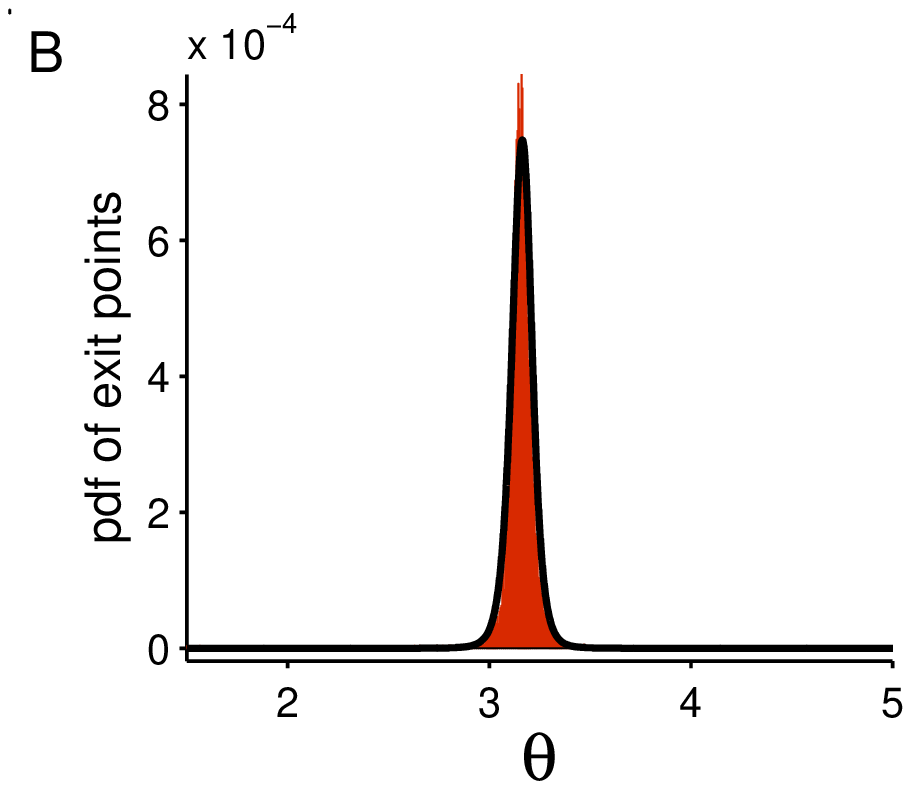}
\caption{{\bf Density of exit points}.(Left): Exit points are marked red. (Right)
Histogram of exit points (red) and the density $P_{\alpha}(\theta)$ from (\ref{proba})
(black). Simulation parameters are $\alpha = -0.9, \omega=10,
\varepsilon=0.005$. Number of runs = 50000.}
\label{compare}
\end{figure}

\begin{figure}[ht!]
\centering
\includegraphics[width=0.5\textwidth]{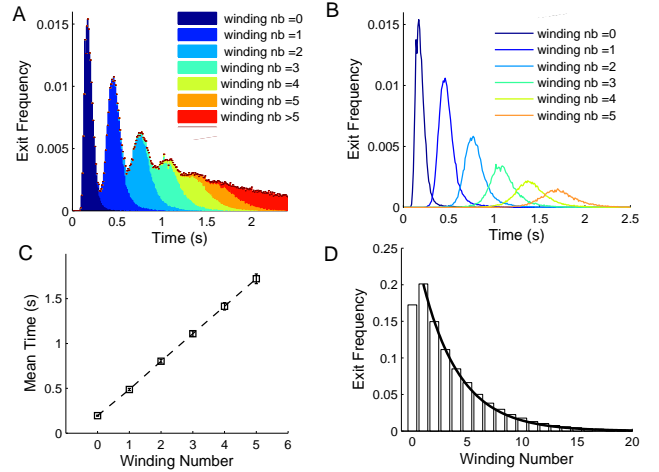}
\caption{\textbf{Statistic of exit trajectories and winding. A:} The different colors represent exit frequencies with
different winding numbers. \textbf{B:} Exit frequencies conditioned on winding number
(0 to 5 turns). \textbf{C:} Mean of exit times in histogram B vs winding number.
\textbf{D:} Histogram of winding numbers in {\bf A}. The exponential function
$f(n)=p(1-p)^{n-1}$ (solid line) approximates the decay rate. Here $p$, the ratio
of frequencies of 1 to 2 turns, is the probability to exit without making a turn.}
\label{decomposed}
\end{figure}

To find the boundary layer equation, we introduce the stretched variable
$\zeta=\rho/\sqrt{\eps}$ and define $q_{\ds\eps}(\x)= Q(\zeta,s,\eps)$.
Expanding all functions in (\ref{WKBMD1}) in powers of $\eps$ we find
\beq
Q(\zeta,s,\eps)\sim Q^0(\zeta,s)+\sqrt{\eps}Q^1(\zeta,s)+\cdots,\label{Qexp17}
\eeq
and we obtain the boundary layer equation
\beq
\sigma(s)\frac{\p ^2Q^0(\zeta,s)}{\p \zeta^2}&-&
\zeta\left[b^0(s)+2\sigma(s)\phi(s)\right] \frac{\p Q^0(\zeta,s)}{\p
\zeta}\nonumber \\
&-&B(s)\frac{\p  Q^0(\zeta,s)}{\p  s} =0,\label{ble17}
\eeq
where $\sigma(s)= (\sigma(0,s)\n.\n)$. The solution that satisfies the boundary and matching conditions
$Q^0(0,s)=0,\quad\lim_{\zeta\to-\infty} Q^0(\zeta,s )=1$
is given by
\begin{align}
Q^0(\zeta,s)=-\sqrt{\frac2\pi}\int\limits_0^{\xi(s)\zeta}e^{-z^2/2}\,dz,\label{1stef}
\end{align}
where $\xi(s)$ is the $S$-periodic solution of the Bernoulli equation
\begin{align}
\sigma(s)\xi^3(s)+[b^0(s)+2\sigma(s)\phi(s)]\xi(s)+B(s)\xi'(s)=0.\label{Berneq2}
\end{align}
\indent{\em The principal eigenvalue.}
The principal eigenvalue is the reciprocal of the MFPT $\bar{\tau_\eps}$ of
$\x_\eps(t)$ to $\partial D$, that is, $\lambda_0\sim1/\bar{\tau_\eps}$.
When $\lambda_0$ is exponentially small, it determines the slow decay of the survival
probability. For small $\eps$ the MFPT $\bar\tau_{\eps}$ is given by
\beq\label{tau}
\bar\tau_{\eps}=\frac{\pi^{3/2}\sqrt{2\eps}}{\ds\,\sqrt{\hbox{det}
\mb{H}}\int_0^{2 \pi}K_0(0,s)\xi(s)\,ds} \exp\left\{\frac{\hat\psi}{\eps}\right\}
\eeq
where $\mb{H}$ the solution of the Riccati equation
$2\mb{H}\mb{\sigma}(\x_0) \mb{H}+\mb{H}\B+\B^T\mb{H}=\mb{0}$,
the function $\xi(s)$ is defined in (\ref{Berneq2}), and \cite{Matkowsky}
\beq
K_0(0,s)=\frac{\xi(s)}{B(s)}\label{K0s}.
\eeq
{\em The full spectrum.}
To compute the higher order eigenfunctions, we set $\eta=\xi(s)\zeta$ and obtain the
boundary layer equations
\beq
\frac{\p ^2\tilde Q^0(\eta,s)}{\p \eta^2}&+& \eta \frac{\p\tilde Q^0(\eta,s)}{\p
\eta}+\frac{B(s)}{\sigma(s)\xi^2(s)}\frac{\p  \tilde Q^0(\eta,s)}{\p  s} \nonumber\\
&=&-\frac{\lambda}{\sigma(s)\xi^2(s)} \tilde Q^0(\eta,s).
\eeq
Separating $\tilde Q^0(\eta,s)=R(\eta)T(s)$, we obtain for the even function $R(\eta)$
the eigenvalue problem
\beqq
R''(\eta)+\eta R'(\eta)+\mu R(\eta)=0, R(0)=0, \lim_{\eta\to-\infty}
R(\eta)=0,\label{blevp}
\eeqq
where $\mu$ is the separation constant and
the eigenvalues  are $\mu_n=2n,\ (n=1,2,\ldots)$ with the
eigenfunctions
$R_n(\eta)=\exp\left\{-\frac{\eta^2}{2}\right\}H_{2n+1}
\left(\frac{\eta}{\sqrt{2}}\right)\label{Rn},$
where $H_{2n+1}(x)$ are the  Hermite polynomials of odd orders \cite{Abramowitz}.  The
function $T(s)$ is given by
\begin{align}
T(s)=\exp\left\{-\lambda\int_0^s\frac{ds'}{B(s')}+
2n\int_0^s\frac{\sigma(s')\xi^2(s')}{B(s')}\,ds'\right\}.
\end{align}
The $2\pi$-period gives for $n=1,\ldots$, the eigenvalues
\begin{align}
\lambda_{m,n}=&\,\left[\frac{n}{\pi}\int_0^{2
\pi}\frac{\sigma(s)\xi^2(s)}{B(s)}\,ds+mi\right]\tilde\omega.
\label{moi}\\
\tilde \omega=&\,\frac{2\pi}{\ds\int_0^{2\pi}\frac{ds}{B(s)}}.\nonumber
\end{align}
Thus the expressions (\ref{tau}), and (\ref{moi}) define the full spectrum as
\beqq
Sp(L)=\left\{\lambda_0[1+O(\eps)],\bigcup_{\stackrel{n\geq 0}{
m=\pm1,\pm2,\ldots}}\lambda_{m,n}[1+O(\eps)]\right\}.
\eeqq
{\bf A generic model.}
To illustrate the theory, we compute the exit distribution of {a generic stochastic system when} the focus
$\x_0$ is moved close to the limit cycle $\p D$.  Oscillations in the probability density of the exit time (\ref{pdf}) become discernible (see Fig.\ref{f:trajectories}B), as shown in the following example: the complex plane Hopf-system
\beq
\mb{b}(z)=z(-1+|z|^2+i\omega),
\eeq
where $\omega >0$ is the angular velocity of the field, has a focus at the origin and
its unstable limit cycle is the unit circle. The M\"obius transformation
\beq
\Phi_\alpha (z) = \dfrac{z-\alpha}{1-\alpha z},\quad 0<\alpha<1,
\eeq
moves the focus to $\x_0=(-\alpha,0)$ and leaves the limit
cycle invariant. The resulting field is given by
\beq\label{field}
\mb{b}_{\alpha}(z)= \dfrac{(z+ \alpha)(1+\alpha z)}{(1-\alpha^2)}\!\left(\!-1\! +
\left|\dfrac{z+\alpha}{1+\alpha z}\right|^2\! +\! i\omega \right)\!. 
\eeq
The decomposition (\ref{arhos}) is given by
\beqq \label{coor}
\mb{b}_{\alpha}(\rho, \theta) = -\rho [b^0_{\alpha} (\theta) + O(\rho^2)]\mb{\nu} +
b_{\alpha}^*(\rho,\theta)\Tt,
\eeqq
where $\Tt$ and $\mb{\nu}$ are the unit tangent and outer-normal to
$\partial D$ at $(0,\theta)$, respectively. The components $b_{\alpha}^0 (\theta)$ and
$b_{\alpha}^*(\rho,\theta)$ are given by
\beq
b^0_{\alpha} (\theta) &=& \dfrac{2(1-\alpha^2 -  \omega \alpha \sin
(\theta))}{1-\alpha^2}+O(\rho),\label{b}\\
b_{\alpha}^* (0,\theta) &=& \dfrac{\omega}{1-\alpha^2}(1+ 2 \alpha
\cos (\theta)+ \alpha^2)+O(\rho).\nonumber
\eeq
The long-time probability density $P_{\alpha}(\theta)$ of exit points
on $\p D$ is determined by the principal eigenfunction (\ref{1stef}) and the
$2\pi$-periodic solution $\xi_{\alpha}(s)$ of the Bernoulli equation (\ref{Berneq2}).
In the case at hand (\ref{Berneq2}) takes
the form
\begin{align*}
-\sigma(\theta)\xi_{\alpha}^3(\theta)+[b_{\alpha}^0(\theta)]\xi_{\alpha}(\theta)+
B_{\alpha}(\theta)\xi_{\alpha}'(\theta)=0
\end{align*}
and $\sigma(\theta)=\mb{\sigma}(0,\theta)\mb{\nu}(0,\theta)\cdot\mb{\nu}(0,\theta)$.
The pdf $P_{\alpha}(\theta)$ is given by \cite{Matkowsky}, \cite[eq.(10.127)]{DSP}
\beq
P_{\alpha}(\theta) =
\frac{\ds\frac{\xi_{\alpha}^2(\theta)\sigma(\theta)}{b_{\alpha}^* (0,\theta)}}
{\ds{\int_0^{2\pi}}\ds\frac{\xi_{\alpha}^2(s)\sigma(s)}{b_{\alpha}^*
(0,s)}\,ds},\label{pdfexit}
\eeq
which leads after computations to
\beq
P_{\alpha}(\theta)= \frac{\left( 1+2\alpha \cos \theta +\alpha^2\right)^{-3}}
{\int_{-\pi}^{\pi} \left( 1+ 2\alpha \cos s +\alpha^2 \right)^{-3}\,ds}.\label{proba}
\eeq
As $\alpha\to1$, the density $P_{\alpha}(\theta)$ concentrates at $\theta=\pi$ (Fig.\ref{compare}). Brownian simulations show that the distribution of exit points on $\p D$ is almost zero, except for a small interval $R(\alpha)$ centered at the boundary point closest to focus.
This result shows that exit occurs only in a small arc of the boundary. When a
trajectory fails to hit the boundary while in the neighborhood of this arc, it has to
wind around the focus and return to $R(\alpha)$. In the Brownian dynamics simulations of the noisy Hopf system
we followed the history of each trajectory and evaluated its contribution to the exit time distribution by recording its winding number prior to exit. The winding renders the average lengths of exiting trajectories quantized by the winding numbers around the focus, as confirmed in the empirical statistics presented in  Fig.\ref{decomposed}.

To evaluate the MFPT (\ref{tau}), we find that $\mb{H}_{\alpha}=\mb{I}$,
$\hat\psi_{\alpha}= \psi_{\alpha}(-1) = \frac{1}{2} (1-\alpha)^2$, and direct
integration gives
\begin{align}
&\,\int\limits_0^{2 \pi}K_0(0,s)\xi_{\alpha}(s)\,ds =\frac{4\pi \left(
{\alpha}^{4}+4\,{\alpha}^{2}+1 \right)  }{C(\omega)(1+\alpha^2)},
\label{den}\\
&\,C(\omega)=\frac{3\omega}{8} -
\frac{8/\omega}{1+(4/\omega)^2}+\frac{4/\omega}{4+(4/\omega)^2)}>0.\nonumber
\end{align}
It follows that for $(1-\alpha)^2/2\eps=O(1)$,
\begin{align*}
\bar\tau_{\eps}\sim&\, \dfrac{C(\omega)\sqrt{2\pi
\eps}(1+\alpha)^2}{4(1+4\alpha^2+\alpha^4)}\exp\left\{\frac{\hat\psi_{\alpha}}{\eps}\right\}
\nonumber\\
\sim&\frac{C(\omega)\sqrt{2
\pi \eps}}{6} \exp\left\{\frac{(1-\alpha)^2}{2\eps}\right\}=O(1),
\end{align*}
so it is not exponentially long in $\eps^{-1}$ as in Kramers' and the classical
exit problems.\\
{\em The second eigenvalue}.
To determine the second eigenvalue from (\ref{moi}), we note that here
$b_{\alpha}^* (0,\theta) = \omega, b^0_{0} (\theta)= 2$, so $\tilde \omega =\omega$
and $\omega_1(\alpha)=4$. We conclude with the surprising result that the period of the peak in $f_{etd}(t)$ is $2\pi/\omega$, where $\omega$ is precisely the frequency at the focus point. Furthermore, using
Brownian simulations, we found that $f_{etd}(t)$ can be well approximated by the sum of the first two exponentials
$\tilde f_{etd}(t)=C_0e^{- \lambda_0 t}+ C_1e^{-\omega_{1}t}\cos (\omega t+\phi)$,
where $\lambda_0$ is the principal eigenvalue and $C_0,C_1,\phi$ are constants.\\
{\bf Discussion and conclusion}\\
 We have demonstrated that the oscillation of the exit distribution is an intrinsic property of dynamical systems driven by noise. As the focus is moved toward the limit cycle, the oscillations become discernible. The frequency of the
peaks is the oscillation frequency of the deterministic dynamical system near the attractor. This is surprising, because the oscillation is not affected by the noise amplitude.  As the focus moves toward the limit cycle, the first eigenvalue becomes of order 1, but also the second eigenvalue is of order one and does not change. This phenomenon is observable in a class of escape problems from the domain of attraction of a stable focus across the unstable limit cycle bounding the domain. Richer phenomenology should be expected in dimensions higher than two. Note that the peak oscillations are not related to stochastic resonance and can potentially be used to interpret physical and biological escape phenomena from their physical model.


\end{document}